\documentclass[a4paper]{article}
\usepackage{ISCSLP2022}
\usepackage{amsthm,amsmath,amssymb,graphicx}
\usepackage{subfigure,caption}
\usepackage{url}
\usepackage{booktabs}
\usepackage{multirow}

\title{Deep Learning Based Audio-Visual Multi-Speaker DOA Estimation Using Permutation-Free Loss Function}
\name{Qing Wang$^{1}$, Hang Chen$^{1}$, Ya Jiang$^{1}$, Zhe Wang$^{1}$, Yuyang Wang$^{1}$, Jun Du$^{1*}$, Chin-Hui Lee$^{2}$}
\address{
  $^{1}$University of Science and Technology of China, Hefei, China\\
$^{2}$Georgia Institute of Technology, Atlanta, GA. USA}
\email{jundu@ustc.edu.cn}

\begin{document}

\maketitle
\begin{abstract}
  In this paper, we propose a deep learning based multi-speaker direction of arrival (DOA) estimation with audio and visual signals by using permutation-free loss function. We first collect a data set for multi-modal sound source localization (SSL) where both audio and visual signals are recorded in real-life home TV scenarios. Then we propose a novel spatial annotation method to produce the ground truth of DOA for each speaker with the video data by transformation between camera coordinate and pixel coordinate according to the pin-hole camera model. With spatial location information served as another input along with acoustic feature, multi-speaker DOA estimation could be solved as a classification task of active speaker detection. Label permutation problem in multi-speaker related tasks will be addressed since the locations of each speaker are used as input. Experiments conducted on both simulated data and real data show that the proposed audio-visual DOA estimation model outperforms audio-only DOA estimation model by a large margin.
\end{abstract}

\section{Introduction}
\label{sec:intro}

Sound source localization (SSL) aims to estimate the position of single or multiple sound sources relative to the position of the recording microphone array. In most cases, we are interested in direction of arrival (DOA) for each sound source, hence most of the SSL methods focus on azimuth and elevation angles estimation. Effective SSL is of great importance in many applications including automatic speech recognition (ASR) \cite{lee2016dnn}, tele-conferencing \cite{zhang2008maximum}, robot audition \cite{valin2003robust,keyrouz2014advanced}, and hearing aids \cite{farmani2017informed}. 

Many previous studies about SSL pay more attention to audio modality alone.  Conventional SSL methods, such as generalized cross-correlation with phase transform (GCC-PHAT) \cite{knapp1976generalized}, steered response power with phase transform (SRP-PHAT) \cite{do2007real}, estimation of signal parameters via rotational invariance technique (ESPRIT) \cite{roy1989esprit}, and multiple signal classification (MUSIC) \cite{schmidt1986multiple}, were based on signal processing techniques and usually performed poorly in noisy and reverberant environments. Deep neural network (DNN)-based SSL methods have been proposed in recent years and proven to outperform conventional SSL methods due to their strong regression capability \cite{liu2018direction}. Grumiaux \cite{grumiaux2021review} provided a thorough survey of the audio SSL literature based on deep learning techniques. The output strategy for DOA estimation can be divided into two categories: classification and regression. Convolutional recurrent neural networks (CRNN) were proposed for DOA estimation of multiple sources by using a classification strategy in \cite{adavanne2018direction,perotin2019crnn}. Some other works \cite{he2021neural,wang2021four} tried to solve the SSL problem as a regression task by directly estimating either Cartesian coordinates or spherical coordinates. Tang \cite{tang2019regression} demonstrated that regression model achieved better performance over classification model. 

By hearing and seeing, human brain is able to perceive surroundings and extract complementary information. Intelligent devices equipped with audio-visual sensors are supposed to achieve similar goals. Fusion of audio and video modalities has shown promising results in many areas, e.g. acoustic scene classification \cite{wang2021curated}, speech enhancement \cite{chen2021correlating}, and active speaker detection \cite{kopuklu2021design}. The literature on audio-visual localization is sparse compared to the large number of studies for sound source localization \cite{grumiaux2021review}. Most of these works \cite{tian2018audio,senocak2018learning,sanguineti2021audio} mainly focused on localizing sound sources in video clips rather than estimating DOA of sound sources. In \cite{qian2021multi}, the authors first proposed a deep neural network (DNN) architecture for audio-visual multi-speaker DOA estimation by simulating visual features. Promising results were observed in \cite{qian2021multi} when at most two speakers existed however the performance of localizing more than two speakers remained unknown. Berghi \cite{berghi2022visually} proposed a teacher-student model to perform active speaker detection and localization with the `teacher' network generating pseudo-labels and the `student' network localizing speakers.

Most of previous works only consider localizing one or two concurrent speakers and the existing audio-visual datasets \cite{lathoud2004av16,he2018deep,berghi2022visually} are of limited size. In this paper, we propose a novel audio-visual DOA estimation approach for multi-speaker scenario based on the MISP2021-AVSR corpus \cite{chen2022misp}, a large-scale audio-visual Chinese conversational corpus which contains 141 hours of audio and video data with at most six concurrent speakers. To avoid expensive and time-consuming cost of manual annotation, we propose to produce the ground truth of DOA for each speaker based on the video data and camera calibration. Then we solve the multi-speaker DOA estimation problem as active speaker detection with the ground truth of DOA served as complementary input to acoustic feature. Label permutation problem in multi-speaker related tasks will be addressed since the locations of each speaker are used as input.


\section{Proposed Method}
\label{sec:pro}
In this section, we describe our proposed approach for multi-speaker DOA estimation using audio and video data. Real data is recorded in the home TV scenario with several people sitting and chatting in Chinese. In home TV scenario, people are always sitting, so our study is focused on estimating the azimuth angle only. Firstly, we introduce how to generate DOA labels with video clips. Then we describe the proposed multi-modal DOA (MDOA) and audio-only DOA (ADOA) estimation models for multi-speaker situation.

\subsection{Spatial Annotation Method}
\begin{figure}[t]
	\centering
	\includegraphics[width=2.4in]{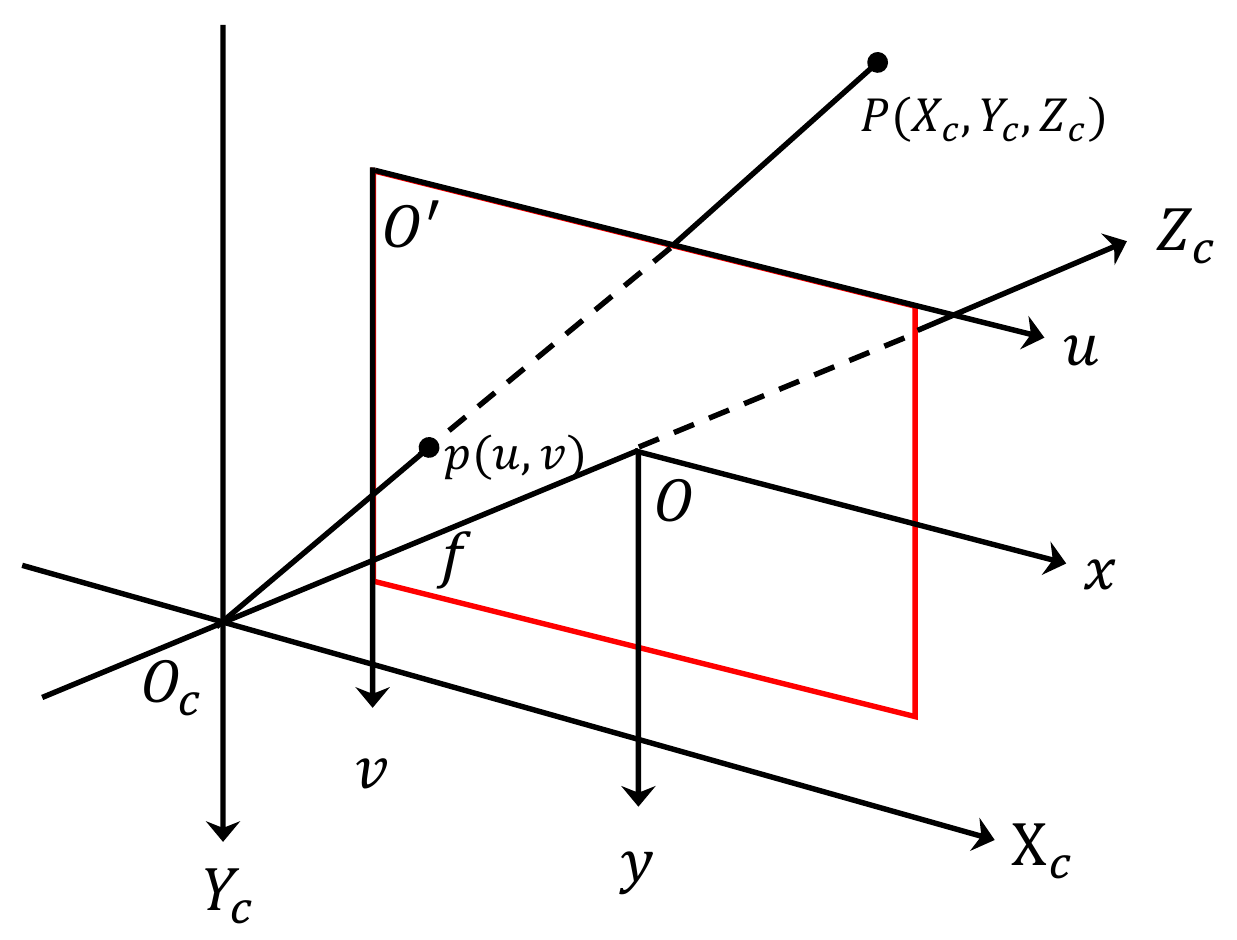}
	\caption{Transformation from pixel coordinate to camera coordinate. $(X_C,Y_C,Z_C)$: camera coordinate; $(x,y)$: image coordinate; $(u,v)$: pixel coordinate.}
	\label{fig1}
\end{figure}
It is expensive and time-consuming to annotate DOA labels manually, therefore we propose to produce the ground truth of DOA with the video data by transformation between camera coordinate and pixel coordinate according to the pin-hole camera model \cite{hartley2003multiple}. The linear microphone array is placed on in the horizontal axis $x$ of the camera coordinate system, with its center coinciding with the origin of the coordinate system. Based on this, we transfer the target point in the pixel coordinate to the camera coordinate and use the resulting point as the sound source location.

Several effective techniques were used to detect the head-and-shoulder of target speakers in the video. In particular, we adopted a head-and-shoulder detection model based on yolo-v5\footnote{https://github.com/ultralytics/yolov5} and the Deep SORT algorithm \cite{wojke2017simple} to track and match multiple speakers in the same video clips simultaneously. The average missing rate of head-and-shoulder detection is less than 1\% at the frame level on MISP2021-AVSR dataset. We average the pixel coordinates of the top-left and bottom-right points of the head-and-shoulder detection bounding box, which is considered to be the location of the mouth. 

Figure~\ref{fig1} shows the transformation process from pixel coordinate to camera coordinate. $O_{c}$, $O$ and $O^{'}$ denote the origins of camera coordinate, image coordinate and pixel coordinate, respectively. $Z_{c}$ corresponds to the camera's optical axis and $f$ represents the image distance. Let us define $p(u,v)$ as the pixel coordinate of one target speaker in the image plane. Then the corresponding point in the camera coordinate is denoted as $P(X_{c},Y_{c},Z_{c})$. By using the pin-hole camera model \cite{hartley2003multiple}, point in the image coordinate is expressed as follows:
\begin{equation}
\begin{pmatrix}x\\y\\1 \end{pmatrix}=\frac{1}{Z_c}\begin{pmatrix}f &0 &0 &0 \\0 &f &0 &0 \\ 0 &0 &1 &0 \end{pmatrix} \begin{pmatrix}X_c\\Y_c\\Z_c\\1 \end{pmatrix}
\end{equation}
The affine transformation between image coordinate and pixel coordinate can be written as:
\begin{equation}
\begin{pmatrix}u\\v\\1 \end{pmatrix}=\begin{pmatrix}\frac{1}{dx} &0 &u_0 \\0 &\frac{1}{dy} &v_0 \\ 0 &0 &1 \end{pmatrix} \begin{pmatrix}x\\y\\1 \end{pmatrix}
\end{equation}
where $dx$ and $dy$ represent the physical length of each pixel on the horizontal axis $x$ and the vertical axis $y$, respectively. The coordinates of point $O$ in the pixel coordinate system are represented by $u_0$ and $v_0$. And  $\lambda=\begin{pmatrix}\frac{f}{dx} &0 &u_0 &0 \\0 &\frac{f}{dy} &v_0 &0 \\ 0 &0 &1 &0 \end{pmatrix}$ is the intrinsic parameters obtained by camera calibration. With pixel coordinates $p(u,v)$ and intrinsic parameters $\lambda$, we can get the camera coordinates $P(X_c,Y_c,Z_c)$ as follows:
\begin{equation}
P=\Phi(p,\lambda)
\end{equation}
where $\Phi$ is the transformation function. As we use monocular camera in our study, the value of $Z_c$ is always one. Finally, the angle between the vector $O_cP$ and the axis $O_cX_c$ represents the azimuth angle of one target speaker.

We evaluate the correctness of the spatial annotation algorithm by selecting ten points in the recording room and comparing their oracle angles with the annotated angles. The difference of angles is less than $0.5^{\circ}$, which demonstrates  that our proposed spatial annotation algorithm could accurately generate ground truth of DOAs.

\subsection{Neural Network Architecture}
\begin{figure}[t]
    \centering
    \subfigure[ADOA estimation]{
    \begin{minipage}{3.2cm}
    \centering
    \includegraphics[scale=0.25]{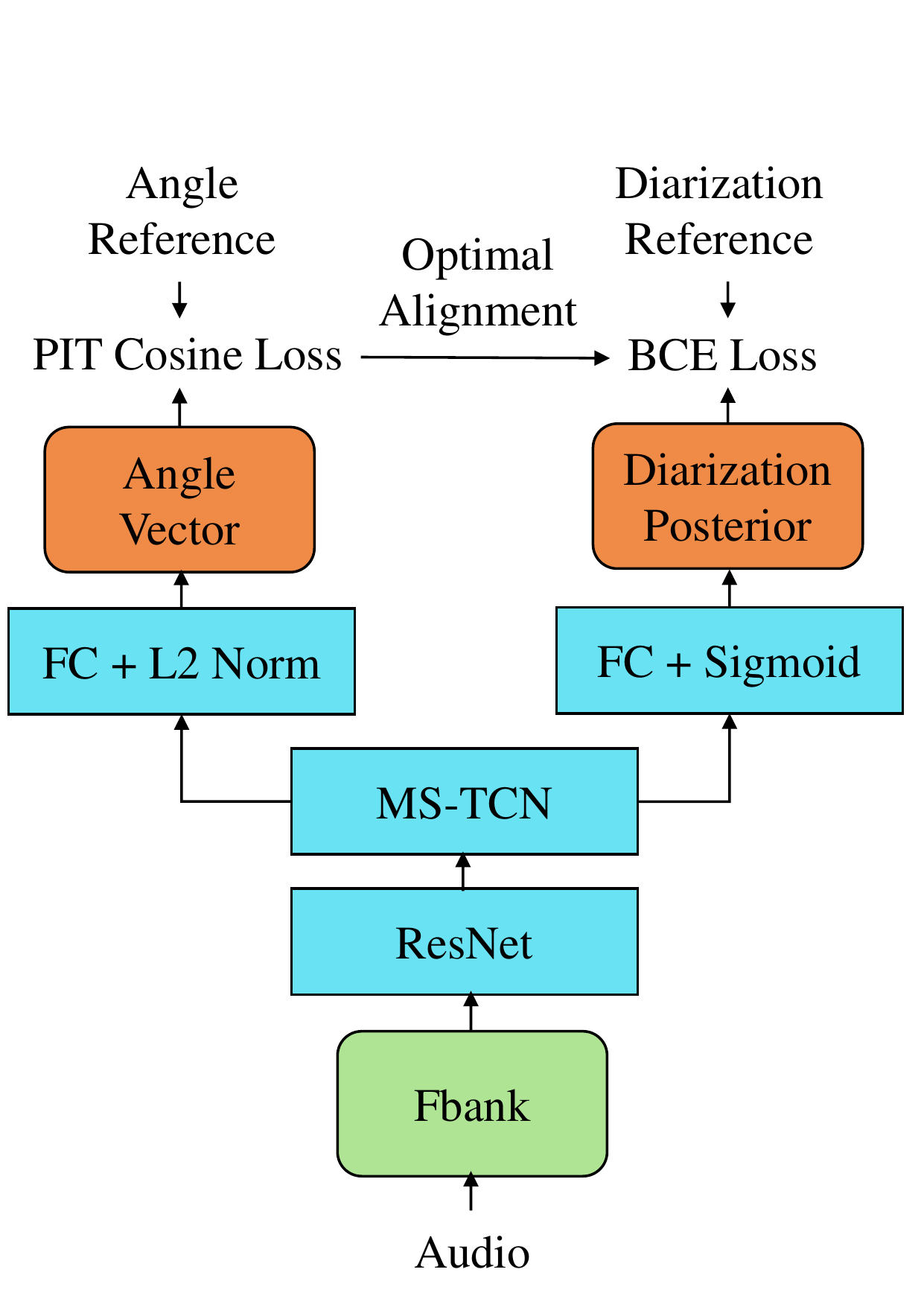}
    \end{minipage}}
    \subfigure[MDOA estimation]{
    \begin{minipage}{3.2cm}
    \centering
    \includegraphics[scale=0.25]{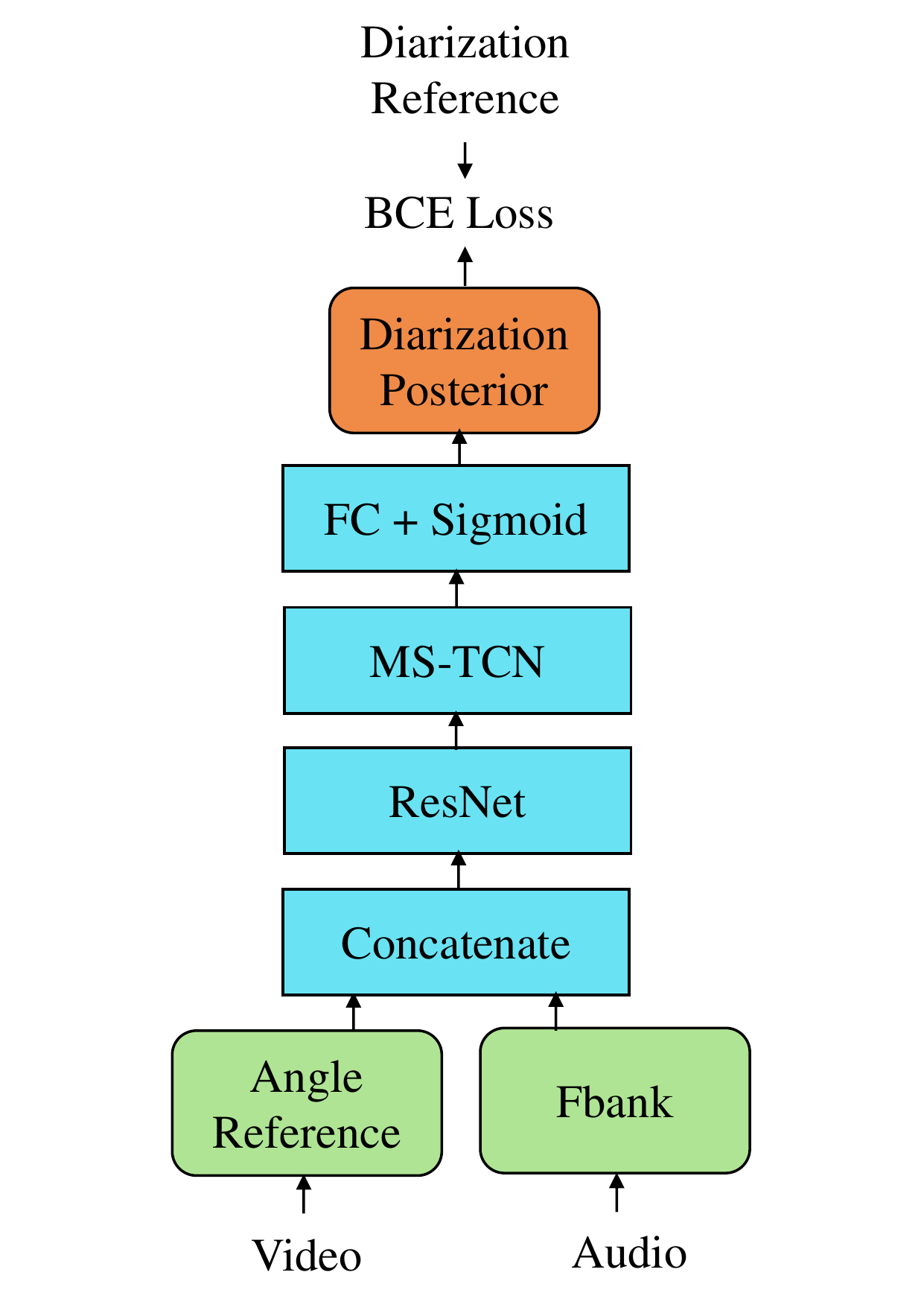}
    \end{minipage}}
    \caption{The overall network architectures of ADOA and MDOA estimation models.}
    \label{fig:framework}
\end{figure}

We design two models to solve ADOA and MDOA estimation for multi-speaker situation. Figure~\ref{fig:framework} illustrates the overall network architectures of ADOA and MDOA estimation models.

For audio data, we extract Mel Filter Bank (Fbank) features, which are then fed into the ResNet \cite{he2016deep} encoder to learn high-level feature representations. ResNet contains several hidden layers with each hidden layer consisting of several blocks. We conduct ablation experiments to find proper ResNet architecture for DOA estimation. Multi-Scale Temporal Convolution Network (MS-TCN) \cite{9053841} is adopted to model the temporal structures within the signal by using several 1D temporal convolutions with different kernel sizes. MS-TCN includes four blocks with each block consisting of two sequential modules. Each module has three branches of temporal convolution with different kernel sizes. The kernel sizes are set to 3, 5, and 7, respectively with the channel number equal to 256. 

There is a big difference of the inputs and outputs between ADOA estimation and MDOA estimation models. For ADOA estimation, only audio data is used as input and there are two branches of fully-connected (FC) layers in the output. We aim to predict the angle vector represented by $cosine$ and $sine$ values of the azimuth angle instead of the angle itself. Hence, L2 normalization is used to make sure that the magnitude of the angle vector is equal to 1. Cosine loss function is adopted to evaluate the angle error between reference azimuth and predicted azimuth. We use permutation invariant training (PIT) strategy to solve the label alignment problem for multiple sound sources. And sigmoid activation is employed to predict diarization posterior of each speaker with binary cross entropy (BCE) loss calculated with the optimal alignment determined by minimizing the Cosine loss. The total loss function for ADOA estimation model is expressed as:
\begin{equation}
E^{\rm{ADOA}} = E_{\rm {cos}}^{\rm{ADOA}}+E_{\rm{bce}}^{\rm{ADOA}}
\end{equation}
where $E_{\rm{cos}}$ and $E_{\rm{bce}}$ are written as:
\begin{equation}
E_{\rm{cos}}^{\rm{ADOA}}=\frac{1}{N}\sum_{n}\frac{\sum_{t}\sum_{s} y_{n,t}^s[1-{\rm{cos}}(\theta_{n,t}^{\phi^*(s)},\hat{\theta}_{n,t}^s)]}{A_n}
\end{equation}
\begin{equation}
E_{\rm{bce}}^{\rm{ADOA}}=\frac{1}{N\times S}\sum_n\frac{\sum_t\sum_sm_{n,t}{\rm{bce}}(y_{n,t}^{\phi^*(s)},\hat{y}_{n,t}^s)}{B_n}
\end{equation}
\begin{equation}
\phi^*=\mathop{\arg\min}\limits_{\phi \in permu(S)}\frac{\sum_{t}\sum_{s} y_{n,t}^s[1-{\rm{cos}}(\theta_{n,t}^{\phi^*(s)},\hat{\theta}_{n,t}^s)]}{A_n}
\end{equation}
where $N$ represents the batch size and $n=1, 2, ...,N$. The two functions  $\rm{cos}(\cdot)$ and $\rm{bce}(\cdot)$ are the Cosine loss and BCE loss, respectively. The frame index $t=1, 2, ..., T$ and the speaker index $s=1, 2, ..., S$ with $S=6$ denoting that the number of output nodes is fixed with 6 in the diarization prediction branch. The angle reference and predicted azimuth angle are represented by $\theta_{n,t}^s$ and $\hat{\theta}_{n,t}^s$ for the $n$-$th$ sample, $t$-$th$ frame and $s$-$th$ speaker, respectively. The diarization reference and predicted diarization posterior are represented by $y_{n,t}^s$ and $\hat{y}_{n,t}^s$, respectively. $A_n=\sum_{t}\sum_{s} y_{n,t}^s$ denotes the total activation number for the $n$-$th$ sample. If the $n$-$th$ sample and the $t$-$th$ frame is silent then $m_{n,t}=0$ otherwise $m_{n,t}=1$, so $B_n=\sum_tm_{n,t}$ represents the number of non-silent frames in the $n$-$th$ sample. And $permu(S)$ is a permutation of $1, 2, ..., S$. Cosine loss is calculated when the speakers are active while BCE loss is calculated for non-silent frames.

It is difficult to predict the azimuth angle for multiple speakers using only audio signals due to the label permutation problem. We propose to solve the multi-speaker DOA estimation task as speaker diarization by feeding the azimuth angles into the model input as shown in Figure~\ref{fig:framework}(b). The MDOA estimation model learns to predict which speaker is active and the corresponding reference azimuth angle of active speaker is then selected as the prediction. It is much easier to choose an angle from a finite candidate set in MDOA than to make a direct prediction of continuous angle in ADOA. BCE loss function is adopted for MDOA estimation model computed on the active frames as follows:
\begin{equation}
E_{\rm{bce}}^{\rm{MDOA}}=\frac{\sum_n\sum_t\sum_sm_{n,t}{\rm{bce}}(y_{n,t}^s,\hat{y}_{n,t}^s)}{S^{\prime}B}
\end{equation}
\begin{equation}
S^{\prime}={\rm{max}}(S_n)
\end{equation}
where $B=\sum_nB_n$ denotes the total number of non-silent frames in the current batch and $S_n$ denotes the number of speakers in the $n$-$th$ sample. The speaker index $s=1, 2, ..., S^{\prime}$ with $S^{\prime}$ denoting the maximum number of speakers among all samples in the batch.

\section{Experimental Results and Analysis}
\begin{table}[t]
	\centering
	\caption{Overview of the real dataset.}
	\label{tab: corpus}
	\footnotesize
	\setlength{\tabcolsep}{6 pt}{
	\begin{tabular}[b]{c|c|c|c|c}
	        \toprule
		Dataset & Training & Validation & Testing & Total\\ \midrule
		Duration (h) & 97.5 & 11.2 & 7.5 & 116.2 \\
		Room & 21 & 5 & 5 & 31 \\
		Speaker & 200 & 21 & 23 & 244 \\ \bottomrule
	\end{tabular}}
\end{table}

\label{sec:exp}
\begin{table*}[htbp]
\centering
\caption{Experimental results on real dataset for ADOA and MDOA estimation models with different parameters of ResNet encoder.}
\label{tab:real}
\begin{tabular}{lllllllll}
\toprule
\multirow{3}{*}{\begin{tabular}[c]{@{}l@{}}Model\end{tabular}} & \multicolumn{4}{c}{ADOA}  & \multicolumn{4}{c}{MDOA}   \\ \cmidrule(r){2-5}\cmidrule(r){6-9} 
& \multicolumn{2}{c}{Validation}  & \multicolumn{2}{c}{Testing} 
& \multicolumn{2}{c}{Validation}  & \multicolumn{2}{c}{Testing} \\\cmidrule(r){2-3}\cmidrule(r){4-5}\cmidrule(r){6-7}\cmidrule(r){8-9}
& \multicolumn{1}{c}{PIMAE $(^{\circ})$} & \multicolumn{1}{c}{ACC} & \multicolumn{1}{c}{PIMAE $(^{\circ})$} & \multicolumn{1}{c}{ACC} & \multicolumn{1}{c}{PIMAE $(^{\circ})$} & \multicolumn{1}{c}{ACC} & \multicolumn{1}{c}{PIMAE $(^{\circ})$} & \multicolumn{1}{c}{ACC} \\ \midrule
Channel32\_Block1  & 21.04  & 0.50  & 22.50  & 0.45                                             & 15.40  & 0.59  & 16.88  & 0.55   \\
Channel32\_Block2  & 21.79  & 0.52  & 22.15  & 0.51                   
                   & 16.13  & 0.58  & 16.75  & 0.54    \\
Channel64\_Block2  & 21.04  & 0.55  & 24.51  & 0.51                    
                   & 15.98  & 0.59  & 16.38  & 0.57   \\
Channel128\_Block2 & 20.80  & 0.53  & 23.83  & 0.45                   
                   & 16.69  & 0.58  & 16.37  & 0.57   \\
Channel128\_Block3 & 27.40  & 0.47  & 30.99  & 0.38                  
                   & 15.10  & 0.62  & 16.96  & 0.56   \\ \bottomrule
\end{tabular}
\end{table*}

\subsection{Experimental Setup}

To evaluate the effectiveness of our proposed MDOA estimation model, we conduct our experiments on a simulated dataset and a real dataset. A python package named pyroomacoustics \cite{scheiblerpyroomacoustics} was adopted to simulate data by generating room impulse responses (RIRs). Audio signals are from TIMIT corpus \cite{garofolo1993darpa} as it consists of a variety of speakers. About 125 hours of data were simulated for the simulated database with 111 hours of data for training and 14 hours of data for testing. The ground truth of DOA for each speaker was calculated with the locations of speakers and microphone array. The simulated audio data is clean without background noise.

Real dataset is recorded in home TV rooms with several people sitting and chatting. Audio data and video data are collected by far-field microphone array and far-field camera, respectively.  The linear microphone array is placed 3-5m away from the speaker, consisting of 6 sample-synchronised omnidirectional microphones. And the distance between adjacent microphones is 35 mm. It is a subset of MISP2021-AVSR corpus \cite{chen2022misp} containing 116 hours of data, half of which is recorded with television on. Details about the real dataset is listed in Table~\ref{tab: corpus}. Note that there is no overlap of speakers and rooms among these three subsets, namely training, validation and testing.

In the test stage, the angle prediction is performed every 100 ms. Assume that the test utterance consists of $L$ frames and the angle prediction of the $l$-$th$ frame is represented by $\hat{\theta}_l$, we select $p_l$ highest peaks of diarization posterior as active speaker prediction with $p_l$ equal to the number of active speakers in the $l$-$th$ frame. And the corresponding DOAs of the $p_l$ speakers are selected as predicted DOAs. 
We use $Permutation$ $Invariant$ $Mean$ $Absolute$ $Error$ (PIMAE) to evaluate the difference between reference DOAs and predicted DOAs. PIMAE is calculated using Hungarian algorithm \cite{kuhn1955hungarian} to find the least angle distance given a set of ground truth angles and its respective predicted angles: 
\begin{equation}
{\rm{PIMAE}} (^{\circ})=\frac{\sum_l\mathcal{H}(\theta_l,{\hat \theta}_l)}{\sum_l p_l}
\end{equation}
where $\mathcal{H}(\cdot)$ represents the Hungarian algorithm; $\theta_l$ and $\hat \theta_l$ denote the reference and predicted angle lists for the $l$-$th$ frame. 
We also adopt $Accuracy$ (ACC) metric \cite{he2021neural} to measure the percentage of correct predictions with a spatial localization error allowance of $20^{\circ}$:
\begin{equation}
{\rm{ACC}}=\frac{\sum_l\sum_{j=1}^{p_l}\boldsymbol{1}_{\rm{<20^{\circ}}}}{\sum_l p_l}
\end{equation}
where $\boldsymbol{1}$ denotes the indicator function and we use the localization error allowance according to the Sound Event Localization and Detection (SELD) task of Detection and Classification of Acoustic Scenes and Events (DCASE) 2021 Challenge\footnote{https://dcase.community/challenge2021/task-sound-event-localization-and-detection}.

We extract 64-dimensional Fbank features for audio data. For ADOA model, the angle vector and diarization posterior in the output layer are set with the shape of $(6,2)$ and $(6)$, respectively. If there are less than six people in one sample, the ground truth of angle and diarization will be padded with zero. For MDOA model, the shape of the diarization posterior in the output layer is determined by $S^{\prime}$, the maximum number of speakers among all samples in the batch, with the angle reference padded with zero when less than $S^{\prime}$ speakers are active in one sample.

\subsection{Experimental Results}
Table~\ref{tab:real} lists the experimental results for ADOA and MDOA estimation models on real dataset with different parameters of ResNet encoder. Television background noises exist in audio signals as interferences. ResNet contains four hidden layers, and the number of channels in each layer is doubled progressively. The term ``Channel32'' denotes that the number of channels in the first hidden layer is set to 32 while the term ``Block1'' denotes that the number of blocks in each hidden layer is set to 1. As shown in Table~\ref{tab:real}, the MDOA estimation method outperforms the ADOA estimation method in all parameter configurations, which demonstrates that with spatial locations served as complementary information, it is more accurate to predict active speakers and their DOAs. The performance of ADOA estimation model varies greatly with different encoder parameters while the MDOA estimation model achieves similar and stable performances when using different number of channels and blocks, which proves the robustness of the MDOA estimation method. Take the third row of ``Channel64\_Block2'' for example, the PIMAE decreases from 24.51$^{\circ}$ to 16.38$^{\circ}$ and the ACC increases from 0.51 to 0.57, achieving a relative 33\% decrease in PIMAE and a relative 12\% increase in ACC. Rather than make a direct prediction of continuous angle, it is easier to choose the correct azimuth angle from a candidate set with video data providing spatial information.

We list the experimental results on simulated dataset in Table~\ref{tab:sim}. The parameter configuration used for ResNet encoder is ``Channel64\_Block2''. Much better results are achieved on simulated dataset than real dataset. This is because that audio signals in Simulated Dataset are not corrupted with noise. For simulated data, the MDOA estimation model achieves 5.77$^{\circ}$ for PIMAE and 0.90 for ACC, yielding a relative 65\% decrease in PIMAE and a relative 27\% increase in ACC compared with ADOA estimation model.

\begin{table}[!htbp]
\footnotesize
\centering
\caption{Experimental results on simulated dataset for ADOA and MDOA estimation models.}
\label{tab:sim}
\begin{tabular}{lclcl}
\toprule
\multirow{2}{*}{Model} & \multicolumn{2}{c}{ADOA}                            & \multicolumn{2}{c}{MDOA} \\ \cmidrule(r){2-3}\cmidrule(r){4-5} 
& PIMAE $(^{\circ})$ & \multicolumn{1}{c}{ACC} & PIMAE $(^{\circ})$           & \multicolumn{1}{c}{ACC} \\ \midrule
Channel64\_Block2                     & \multicolumn{1}{l}{16.67} & 0.71                    & \multicolumn{1}{l}{5.77} & 0.90                    \\ \bottomrule
\end{tabular}
\end{table}

\begin{figure}[h]
	\centering
	\includegraphics[width=2.9in]{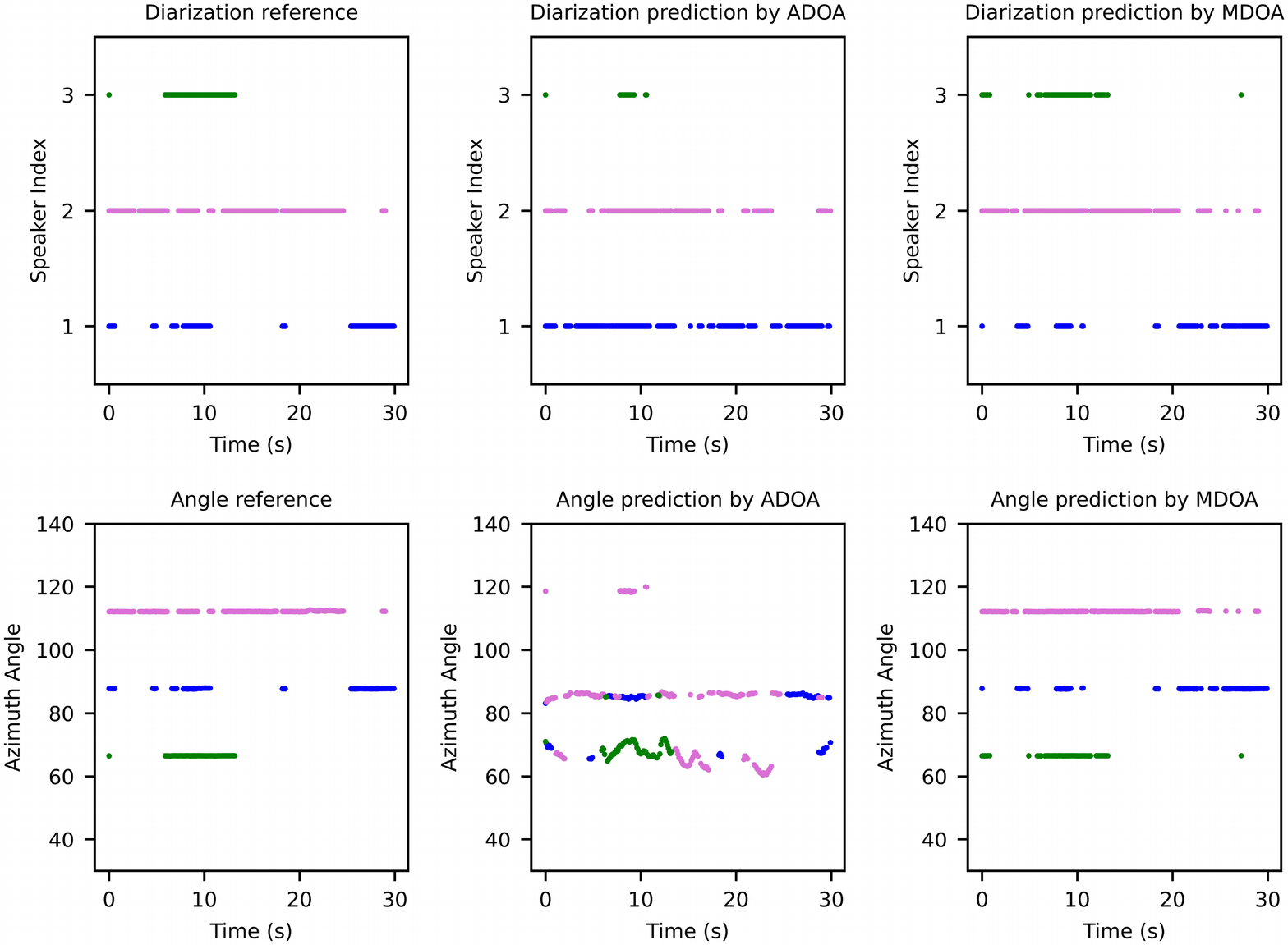}
	\caption{Prediction results for a test sample.}
	\label{demo}
\end{figure}

Figure~\ref{demo} shows a visualization of the prediction results for a test sample. The MDOA model outperforms the ADOA model in both diarization and angle predictions. At about 10 seconds, there are three concurrent speakers, which is difficult to directly predict the continuous angles and large localization distance is got by the ADOA model. However, accurate azimuth angles are predicted by the MDOA model with useful spatial location information of the speakers. From about 10 to 20 seconds, only one speaker (with index 2) is talking and the MDOA model can correctly predict the active speaker and the corresponding angle. Whereas the ADOA model makes a wrong prediction about active speaker at some segments with not consistent angles.  

\section{Conclusion}
\label{sec:con}
In this paper, we propose a deep learning based approach for multi-speaker DOA estimation using permutation-free loss function with audio-visual data. A novel spatial annotation method is adopted to generate the ground truth of DOA for each speaker with the video data according to the pin-hole camera model. By using spatial location information as complementary input, multi-speaker DOA estimation could be solved as a classification task of active speaker detection with permutation-free loss function, which provides a new perspective on multi-modal sound source localization. Experiments on real and simulated datasets demonstrate the superior performance of our proposed model compared to audio-only DOA estimation model in terms of both PIMAE and ACC metrics.

\bibliographystyle{IEEEtran}

\bibliography{mybib}


\end{document}